\documentclass[sigconf]{acmart}

\usepackage{graphicx}
\usepackage{amsmath}

\usepackage{caption}
\usepackage{subfigure}
\usepackage[misc]{ifsym}
\usepackage[normalem]{ulem}
\useunder{\uline}{\ul}{}
\usepackage{threeparttable}
\usepackage{float}

\AtBeginDocument{%
  }

\setcopyright{acmcopyright}
\copyrightyear{2018}
\acmYear{2018}
\acmDOI{XXXXXXX.XXXXXXX}

\acmConference[Conference acronym 'XX]{Make sure to enter the correct
  conference title from your rights confirmation emai}{June 03--05,
  2018}{Woodstock, NY}
\acmPrice{15.00}
\acmISBN{978-1-4503-XXXX-X/18/06}

\begin{document}

\title{Modeling Complex Dependencies for Session-based Recommendations via Graph Neural Networks}

\author{Qian Zhang}
\affiliation{
  \institution{Qilu University of Technology (Shandong Academy of Sciences)}
  \city{Jinan}
  \country{China}
}
\email{qianzhang9706@gmail.com}

\author{Wenpeng Lu}
\authornote{Corresponding Author.}
\affiliation{%
  \institution{Qilu University of Technology (Shandong Academy of Sciences)}
  \city{Jinan}
  \country{China}
}
\email{wenpeng.lu@qlu.edu.cn}

\renewcommand{\shortauthors}{Qian Zhang et al.}

\begin{abstract}

\underline{S}ession-\underline{b}ased \underline{r}ecommendations (SBRs) capture items’ dependencies from the sessions to recommend the next item. In recent years, \underline{G}raph \underline{n}eural \underline{n}etworks (GNN) based SBRs have become the mainstream of SBRs benefited from the superiority of GNN in modeling complex dependencies. Based on a strong assumption of \textit{adjacent dependency}, any two adjacent items in a session are necessarily dependent in most GNN-based SBRs. However, we argue that due to the uncertainty and complexity of user behaviors, adjacency does not necessarily indicate dependency. However, the above assumptions do not always hold in actual recommendation scenarios, so it can easily lead to two drawbacks: (1) \textit{false dependencies} occur in the session because there are adjacent but not really dependent items, and (2) the missing of \textit{true dependencies} occur in the session because there are non-adjacent but actually dependent items. These drawbacks significantly affect item representation learning, degrading the downstream recommendation performance. To address these deficiencies, we propose a novel \underline{r}eview-refined \underline{i}nter-item \underline{g}raph \underline{n}eural \underline{n}etwork (RI-GNN), which utilizes topic information extracted from the reviews of items to improve dependencies between items. Experiments on two public real-world datasets demonstrate that RI-GNN outperforms SOTA methods\footnote{The implementation is available at https://github.com/Nishikata97/RI-GNN.}.
\end{abstract}



\maketitle

\section{Introduction}
In recent years, session-based recommendations (SBRs) have attracted extensive attention \cite{HidasiKBT16,wang2021survey} for its strong capability to capture users' dynamic and short term preference. SBRs recommend the next item to a user by modeling the sequential dependencies over items within  sessions.

Driven by the development of deep learning, many neural network based SBRs have been developed. Among them, recurrent neural network (RNN) and graph neural network (GNN) \cite{wang2021graph} based approaches have shown good performance. RNN-based methods attempt to capture sequential dependencies between items within the sessions, which is based on the assumption that there is a strict chronological order inside the session~\cite{HidasiKBT16}. However, this assumption does not always hold in the real-world scenarios since users' behaviours are usually uncertain and dynamic and thus not all interacted items in one session are sequentially dependent.  
Benefiting from the capability of GNN in learning complex dependencies, many GNN-based SBRs \cite{guo2021sequential,qiu2019rethinking,wang2020global,wu2019session} have been proposed. Leveraging the flexibility of graph structure used in GNN, the problem of strict chronological order confusing RNN-based methods is thus alleviated. 

\begin{figure*}[t]
\centering
\includegraphics[width=0.8\textwidth]{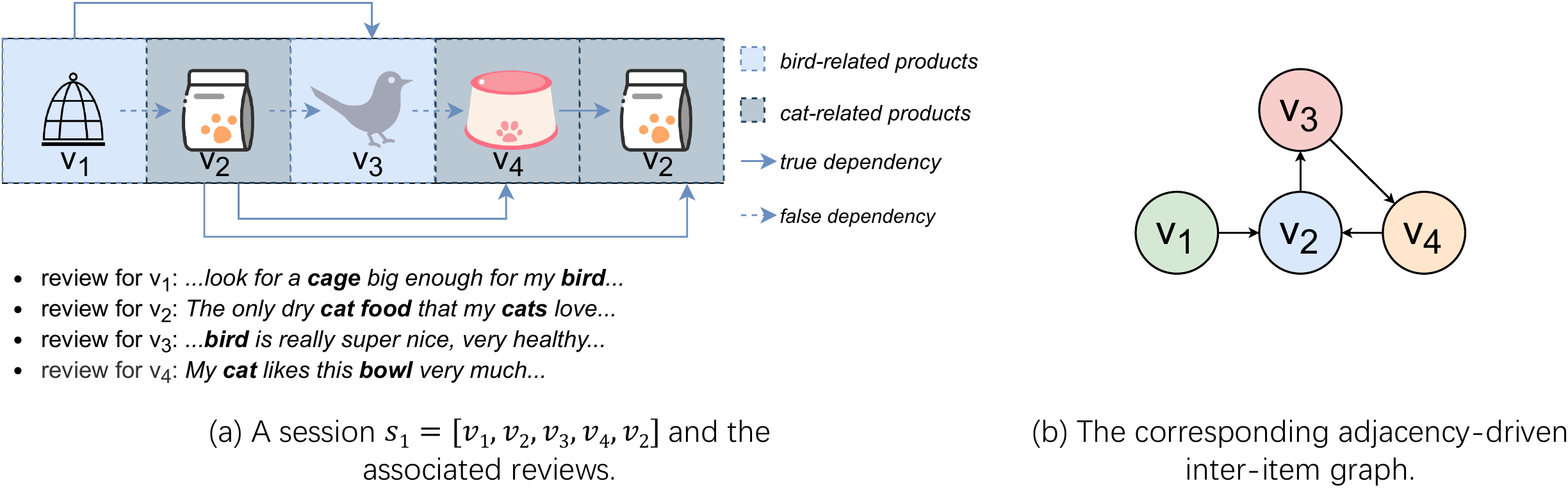}
\caption{A toy example for the construction of adjacency-driven inter-item graph.}
\label{fig1}
\end{figure*}

However, GNN-based SBRs often rely heavily on the strong assumption of adjacent dependency, namely, the adjacent items within one session are necessarily dependent. This is determined by its particular work mechanism. Specifically, most GNN-based SBRs first convert a given session consisting of a sequence of interacted items into a session graph by mapping each item to a node and the adjacency relation between any two items to an edge~\cite{wu2019session} to indicate the dependency between them, as shown in Fig. \ref{fig1}(b). This common practice of constructing session graphs often leads to two significant deficiencies: (1) the introduction of \textbf{false dependencies between adjacent but actually independent items} in a session, e.g., the item $v_1$ (i.e., a bird cage) and item $v_2$ (i.e., cat food) in the session described in Fig. \ref{fig1}(a), and (2) the missing of \textbf{true dependencies between items which are non-adjacent but actually dependent} in a session, e.g., the item $v_1$ and item $v_3$ (i.e., a bird) in the session described in Fig. \ref{fig1}(a). In practice, both false dependencies and true dependencies mentioned above are not uncommon in the real-world cases~\cite{wang2019sequential}. Obviously, these two deficiencies significantly downgrade the accurate learning of inter-item dependencies embedded in session data and thus reduce the performance of the downstream next-item recommendations. Therefore, it is critical to refine the dependencies between items by identifying and keeping all true dependencies while removing the false ones.

In practice, in addition to the session information, the review information associated with items can reveal dependencies between them to some degree. For example, the reviews associated with item $v_1$ and $v_3$ shown in Fig. \ref{fig1} (a) are closely related and fall into the same topic. This actually provides extra information to enable the possibility to refine dependencies between items in sessions. To this end, we propose \underline{r}eview-refined \underline{i}nter-item \underline{g}raph \underline{n}eural \underline{n}etwork (RI-GNN) to address the two deficiencies mentioned above in this paper. By leveraging the topic information from reviews written for items, RI-GNN can not only reduces the false dependencies between adjacent but actually independent items, but also well captures true dependencies between non-adjacent but dependent items which are usually ignored or weaken by existing GNN-based SBRs.

The main contributions of this work are summarized below:
\begin{itemize}
\item We propose and discuss a novel and important research question: \emph{does adjacency necessarily indicate dependency between items in sessions}? We perform a preliminary exploration with the hope of shedding some light in this area.
\item We propose a novel review-refined inter-item graph neural network, called RI-GNN, for session-based recommendations.
To the best of our knowledge, this is the first work for leveraging reviews to enhance the  dependency learning for SBR on anonymous sessions.
\item We propose a novel method for constructing a novel review-refined inter-item graph for each session. In the graph, reviews for items are employed to filter out the false dependencies between some adjacent items, and recall the true dependencies between non-adjacent items missed by existing methods.
\end{itemize}

\section{Related Work}

\subsection{Session-based Recommendation}

Existing methods for SBR can be summarized into: (1) Markov chain-based SBR; (2) RNN-based SBR; (3) Attention-based SBR; and (4) GNN-based SBR.

(1) \emph{Markov chain-based SBR}. Early researches on SBR rely Markov chain to model the short-term dependencies to predict the next item. For example, FPMC \cite{rendle2010factorizing} combined Markov chain and matrix factorization to model sequential behavior between two adjacent items and recommend next item. However, Markov chain-based methods only focus on first-order dependencies between adjacent items, while neglecting the high-order dependencies between long-distance items. 
(2) \emph{RNN-based SBR}. Due to the powerful ability in modeling sequential data, RNN-based methods are applied widely to SBR \cite{HidasiKBT16,li2017neural,liu2018stamp}. GRU4Rec \cite{HidasiKBT16} first applied RNN to SBR, which adopted gated recurrent unit (GRU) to model the dependencies within sessions. However, RNN-based SBR also suffers the similar problem in Markov chain-based methods, which always biases to short-distance items while missing the information from the long-distance items in sessions. 
(3) \emph{Attention-based SBR}. The attention mechanism is applied to further improve SBR \cite{wang2021hierarchical,wang2018attention} by identifying important items within sessions. NARM \cite{li2017neural} first integrated the attention mechanism into SBR to extract the the user's main purpose in the current session. STAMP \cite{liu2018stamp} proposed a short-term memory priority model based on multi-layer perceptron and attention mechanism, which captured the long-term and short-term interests of users. However, the attention mechanism only focuses on few important items that belong to the user's main purpose, while neglecting the other purposes indicated by few inferior items. 
(4) \emph{GNN-based SBR}. Due to the superiority of GNN on modeling transition dependencies between items, GNN-based methods have been applied widely to SBR. SR-GNN \cite{wu2019session} modeled all sessions via directed graphs, utilized GNN to capture the dependencies between items with sessions, and extracted the long-term and short-term interest of users to suggest the next item. FGNN \cite{qiu2019rethinking} proposed to capture the sequence order and latent order in session graph, which devised the weighted attention graph layer to learn item embeddings and session embeddings for more accurate next item recommendation. GCE-GNN \cite{wang2020global} proposed to learn the transitions between items from local and global perspectives simultaneously, so as to make better recommendations by leveraging the information from other sessions. Although GNN-based methods have achieved great success on SBR, all of these approaches construct the session graph according to the adjacent items, which ignore the dependencies from the non-adjacent items. 

\subsection{Review-based Recommendation}
Considering the great value of user reviews on items, some works strive to model reviews to improve the performance of SR \cite{li2019review,zheng2017joint}. DeepCoNN \cite{zheng2017joint} employed two parallel cooperative neural networks to learn user behaviors by exploiting reviews written by the user and learn item properties from the reviews written for the item, then utilized factorization machine to predict item ratings. RNS \cite{li2019review} proposed a review-driven neural sequential recommendation, which learned user's long-term preference according to her historical reviews. Although the existing methods improve the recommendation performance, most of them are devised for the task of rating prediction instead of session-based recommendation. Although RNS is proposed for sequential recommendation, it requires to collect all reviews written by a user according to the user's explicit ID. This means that it is unable to work well on the anonymous session-based recommendation. Neither of the existing works really solve the problem of session-based recommendations based on review information.

\section{Preliminary}

\subsection{Problem Statement}

Let $V=\left\{v_{1}, v_{2}, \ldots, v_{m}\right\}$ represent the whole item set. Each anonymous session $s=[v_{1}, v_{2}, \ldots, v_{n}] (v_{i} \in V)$ is an ordered list of items, where all the items in $s$ are interacted by an anonymous user in a chronological order. We embed each item into the same embedding space and let $\mathbf{h}_{v_{i}} \in \mathbb{R}^{d}$ denote the embedding of item $v_{i}$, where $d$ is the dimensionality. To accurately identify item dependencies within the session, we utilize review information to enhance item representations. Given an item $v_{i}$, all of its reviews are collected to form the review document $D_{i}$, where each word is represented by the corresponding embedding with the dimension $d_{w}$. For the session-based recommendation problem, the goal is to predict the top-$N$ items that the user is most likely to click in the next step.

\subsection{Graph Construction}
In this subsection, we first introduce the \emph{adjaceny-driven inter-item graph }(AIG), which is widely adopted by existing GNN-based approaches \cite{wang2020global,wu2019session}, and then we present a novel graph, i.e., \emph{review-refined inter-item graph }(RIG). RIG is used as an additional graph to complement AIG rather than to replace it by enhancing the learning of true dependencies.  

\textbf{Adjaceny-driven inter-item graph (AIG).} AIG captures important sequential patterns based on pair-wise adjacent items within the current session, which is first proposed by SR-GNN \cite{wu2019session}. AIG converts each session \emph{s} into a directed graph $\mathcal{G}_{s}^{adj}=(\mathcal{V}_{s}, \mathcal{E}_{s}^{adj})$, where $\mathcal{V}_{s}\subseteq V$ denotes the node set, $\mathcal{E}_{s}^{adj}$ denotes the edge set. The weight of each edge is set as the value of the occurrences of the edge divided by the outdegree of the edge's start node. This means that the more frequent occurrences of the edge, the stronger dependency between the items connected by it. The connection matrix $\mathbf{A}_{s} \in \mathbb{R}^{n \times 2 n}$ describes how nodes are connected with each other in the graph, and $\mathbf{A}_{s, i:} \in \mathbb{R}^{1 \times 2 n}$ are the two columns of blocks in $\mathbf{A}_{s}$, corresponding to node $v_{i}$. $ \mathbf{A}_{s} = \mathbf{A}_{s}^{\text {(out) }} \| \mathbf{A}_{s}^{\text {(in) }}$, where $\mathbf{A}_{s}^{\text {(out) }}$ and $\mathbf{A}_{s}^{\text {(in) }}$ are the outgoing and incoming adjacency matrix respectively. $\|$ indicates the concatenation operation, and $n$ is the length of session $s$.

\textbf{Review-refined inter-item graph (RIG).} The aforementioned AIG faces two deficiencies caused by the adjacent but independent items and the nonadjacent but dependent items. To address them, we utilize reviews to refine the session graph, and thus devise the review-refined inter-item graph (RIG). 

\begin{figure*}[t]
\centering
\includegraphics[width=0.8\textwidth]{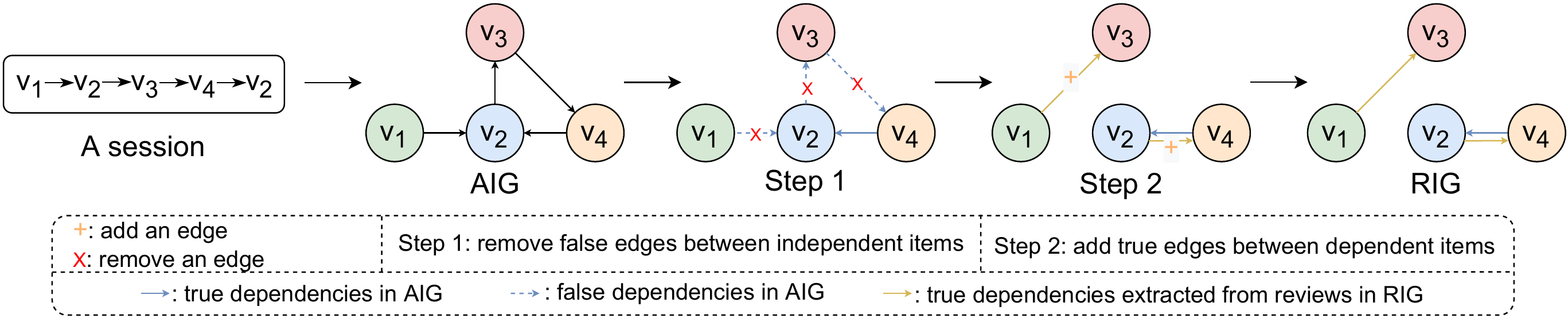}
\caption{Construction of AIG.}
\label{fig3}
\end{figure*}

We convert the session $s$ into a directed graph $\mathcal{G}_{s}^{re}=(\mathcal{V}_{s}, \mathcal{E}_{s}^{re})$, where $\mathcal{V}_{s}\subseteq V$ indicates the node set, $\mathcal{E}_{s}^{re}$ denotes the edge set. As shown in Fig. \ref{fig3}, there are two steps for RIG to recognize the \textit{cross-item dependencies} (i.e., the dependencies between non-adjacent items) to obtain the correct edges. \emph{Firstly}, once AIG is generated, in order to filter out the noise edges and refine the session graph, we only reserve the edges between items sharing the same topic in the AIG, i.e., the same user purpose, and remove the other edges. For recognizing the topic information conveniently, we collect all reviews written for the item into the document set $D$ together, and utilize LDA \cite{blei2003latent} to extract their topics. Once obtaining the topics of each item, we filter out the edges between items that do not belong to the same topic.  \emph{Secondly}, for each ordered pair of nodes $(v_{t_{1}}, v_{t_{2}})$ in the session sequence, we add the directed edge $(v_{t_{1}} \rightarrow v_{t_{2}})$ if item $v_{t_{1}}$ shares the same topic with item $v_{t_{2}}$ and $t_{1} < t_{2}$. This makes the current item directly connect to all the following dependent items in the session for capturing the cross-item dependencies. Finally, we obtain the refined edge set $\mathcal{E}_{s}^{re}$. Similar to AIG, we calculate the connection matrix $\mathbf{B}_{s} \in \mathbb{R}^{n \times 2 n}$ for RIG, where $\mathbf{B}_{s} = \mathbf{B}_{s}^{\text {(out) }} \| \mathbf{B}_{s}^{\text {(in) }}$.

\section{Architecture of RI-GNN Model}

\begin{figure*}[t]
\centering
 \includegraphics[width=0.8\textwidth]{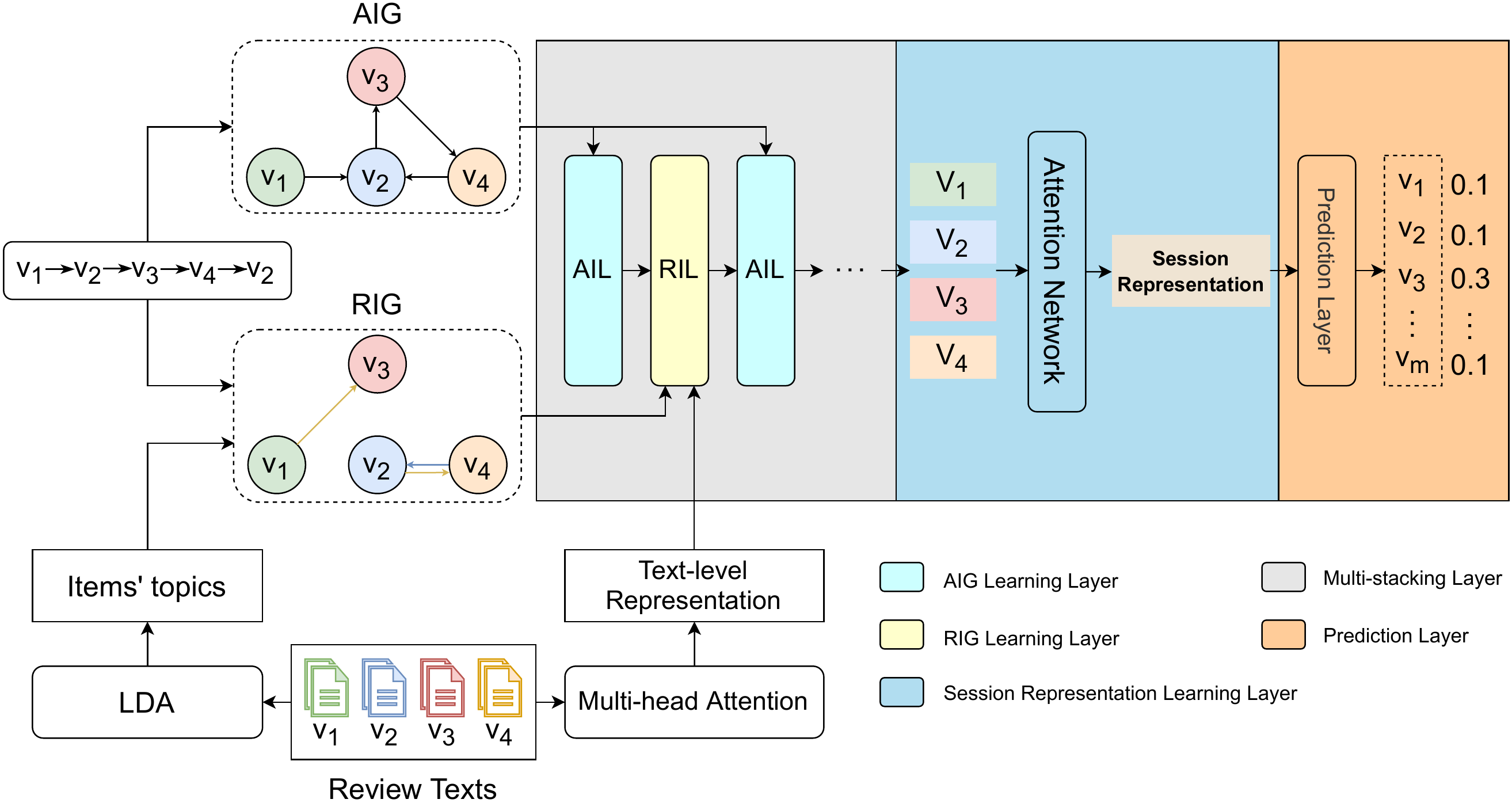}
\caption{Architecture of our proposed RI-GNN model.}
\label{fig4}
\end{figure*}

The architecture of our proposed RI-GNN is shown in Fig. \ref{fig4}, which mainly consists of five components, i.e., \emph{adjacency-driven inter-item graph (AIG) learning layer}, \emph{review-refined inter-item graph (RIG) learning layer}, \emph{multi-stacking layer}, \emph{session representation learning layer} and \emph{prediction layer}.

\subsection{Adjaceny-driven Inter-item Graph Learning Layer (AIL)}
AIL aims to capture the sequential dependencies between items based on AIG within the current session. Next, we will present how to learn the sequential dependencies between adjacent pair-wise items, as follows:

\begin{equation}
\small
\begin{split}
\mathbf{a}_{s, i}^{t} &=\mathbf{A}_{s, i:}^{s}\left[\mathbf h_{{v}_{1}}^{t-1}, \ldots, \mathbf h_{{v}_{n}}^{t-1}\right]^{\top} \mathbf{H}+\mathbf{b_{1}}, \\
\mathbf{z}_{s, i}^{t} &=\sigma\left(\mathbf{W}_{z} \mathbf{a}_{s, i}^{t}+\mathbf{U}_{z} \mathbf h_{{v}_{i}}^{t-1}\right), \\
\mathbf{r}_{s, i}^{t} &=\sigma\left(\mathbf{W}_{r} \mathbf{a}_{s, i}^{t}+\mathbf{U}_{r} \mathbf h_{{v}_{i}}^{t-1}\right), \\
\widetilde{\mathbf{h}}_{v_{i}}^{t} &=\tanh \left(\mathbf{W}_{o} \mathbf{a}_{s, i}^{t}+\mathbf{U}_{o}\left(\mathbf{r}_{s, i}^{t} \odot \mathbf h_{{v}_{i}}^{t-1}\right)\right), \\
\mathbf h_{{v}_{i}}^{t} &=\left(1-\mathbf{z}_{s, i}^{t}\right) \odot \mathbf h_{v_{i}}^{t-1}+\mathbf{z}_{s, i}^{t} \odot \tilde{\mathbf{h}}_{v_{i}}^{t},
\end{split}
\end{equation}
where $\mathbf{a}_{s, i}^{t} \in \mathbb{R}^{2 d}$ is the current state at time step $t$, which aggregates the adjacent items' embeddings for item $v_{i}$ in AIG. $[\mathbf h_{v_{1}}^{t-1}, \ldots, \mathbf h_{v_{n}}^{t-1}]$ is the list of item embeddings in session $s$ at previous time step $t-1$, $\mathbf{A}_{s, i:} \in \mathbb{R}^{1 \times 2 n}$ are the two columns of blocks in $\mathbf{A}_{s}$ corresponding to item $v_{i}^{s}$, $\mathbf{H} \in \mathbb{R}^{d \times 2 d}$, $\mathbf{W}_{z}, \mathbf{W}_{r}, \mathbf{W}_{o} \in \mathbb{R}^{d \times 2d}$, $\mathbf{U}_{z}, \mathbf{U}_{r}, \mathbf{U}_{o} \in \mathbb{R}^{d \times d}$, $\mathbf{b_{1}} \in \mathbb{R}^{2d}$ are trainable parameters, 
$\sigma(\cdot)$ is the sigmoid function, and $\odot$ is the element-wise multiplication operator, $\mathbf{z}_{s, i}^{t} \in \mathbb{R}^{d}$ and $\mathbf{r}_{s, i}^{t} \in \mathbb{R}^{d}$ are the update and reset gates respectively, $\mathbf h_{{v}_{i}}^{t}$ is the final state at the time step $t$. We mark the final representation of item $v_{i}$ in AIL as $\mathbf h_{v_{i}}^{adj}$.

\subsection{Review-refined Inter-item Graph Learning Layer (RIL)}

In order to filter out the noise information and improve the representations of items, we next present how to propagate features on RIG to encode item dependencies from reviews. This layer is built based on the architecture of graph neural network, and we generate attention weights based on the similarity of reviews between items by exploiting the idea of graph attention network \cite{velivckovic2018graph}.

For each item's review document $D_{i}$ obtained from \emph{Section 3.2}, we first convert it into a representation vector $\mathbf{E}_{i} \in \mathbb{R}^{l \times d_{w}}$ through word embeddings. 
In order to better extract item features from the review representation $\mathbf{E}_{i}$, we utilize the self-attention method proposed by Transformer \cite{vaswani2017attention}:
\begin{equation}
\small
\operatorname{Attention}(\mathbf{Q}, \mathbf{K}, \mathbf{V})=\operatorname{softmax}\left(\frac{\mathbf{Q} \mathbf{K}^\mathsf{T}}{\sqrt{d}}\right) \mathbf{V},
\end{equation}
where $\mathbf{Q}$ is the queries, $\mathbf{K}$ is the keys, $\mathbf{V}$ is the values and $\sqrt{d}$ is the scale factor. 
We adopt multi-head attention to enable the model to jointly focus on information from different representation subspaces from different positions. For item $v_{i}$, the detailed operations are described as below:
\begin{equation}
\small
\begin{split}
&\text {head}_{\textit{k}} =\operatorname{Attention}\left(\mathbf{E}_{i} \mathbf{W}_{k}^\mathbf{Q}, \mathbf{E}_{i} \mathbf{W}_{k}^\mathbf{K}, \mathbf{E}_{i} \mathbf{W}_{k}^\mathbf{V}\right), \\
\mathbf{r}_{i} =& \text {MultiHead}(\mathbf{E_{i}}) \left.=\text { Concat(head}_{\textit{1}}, \ldots, \text {head}_{\textit{h}}\right) \mathbf{W}_{1},
\end{split}
\end{equation}
where $\mathbf{r}_{i}$ is the review representation of item $v_{i}$ extracted by the multi-head attention, $\mathbf{W^{Q}} \in \mathbb{R}^{d_{w} \times d_{q}}$, $\mathbf{W^{K}} \in \mathbb{R}^{d_{w} \times d_{k}}$, $\mathbf{W^{V}} \in \mathbb{R}^{d_{w} \times d_{v}}$, and $\mathbf{W_{1}} \in \mathbb{R}^{{h}d_{v} \times d_{w}}$ are the learnable parameters. In our experiments, we set the number of parallel attention heads $h$ to 3, set the dimensions of $d_{q}$, $d_{k}$, $d_{v}$, $d_{w}$ to 100, 100, 100, 300.

In order to distinguish the importance of neighbor items for obtaining the representation of current item, we adopt attention mechanism and calculate the attention weight by cosine similarity:
\begin{equation}
\small
\begin{split}
\pi\left(v_{i}, v_{j}\right)=\left\{\begin{aligned}
\operatorname{sim}\left(\mathbf{r}_{i}, \mathbf{r}_{j}\right), & \text { if } TP_{i} = TP_{j} \\
0, & \text{ if } TP_{i} \neq TP_{j}
\end{aligned}\right.,
\end{split}
\end{equation}
where $\pi\left(v_{i}, v_{j}\right)$ estimates the importance weight of different neighbor items, $\operatorname{sim()}$ is the cosine similarity function, ${\mathbf{r}}_{i}$ and ${\mathbf{r}}_{j}$ are multi-head review representation of item $v_{i}$ and item $v_{j}$ respectively, $TP_{i}$ and $TP_{j}$ are the topics of item $v_{i}$ and item $v_{j}$ respectively. Next, we can obtain the final item representation by the linear combination of neighbor items:
\begin{equation}
\small
\begin{split}
\hat{\pi}(v_{i}, v_{j}) &=\frac{\exp \left(\pi\left(v_{i}, v_{j}\right)\right)}{\sum_{v_{k} \in \mathcal{N}_{v_{i}}^{re}} \exp \left(\pi\left(v_{i}, v_{k}\right)\right)}, \quad
\mathbf{h}_{v_{i}}^{re} =\sum_{v_{j} \in \mathcal{N}_{v_{i}}^{re}} \hat{\pi}\left(v_{i}, v_{j}\right) \mathbf{h}_{v_{j}},
\end{split}
\end{equation}
where $\hat{\pi}(v_{i}, v_{j})$ is attention coefficient normalized by softmax, which means the different contribution of neighbor item $v_{j}$ to the current item $v_{i}$. $\mathcal{N}_{v_{i}}^{re}$ is the neighbor set of item $v_{i}$ in the \emph{RIG}, $\mathbf{h}_{v_{j}}$ is the representation of the neighbor item $v_{j}$ of item $v_{i}$, $\mathbf{h}_{v_{i}}^{re}$ is the final representation of item $v_{i}$ in the RIL.

\subsection{Multi-stacking Layer}
In order to fully capture the deep dependencies between items, inspired by the work of Chen et al. \cite{chen2020handling}, we stack multiple AIL and RIL layers, which can capture the complex dependencies (i.e., both adjacent-item and cross-item) within the session, described as below:

\begin{equation}
\small
\mathbf{h}_{0, v_{i}}^{adj} \rightarrow \mathbf{h}_{1, v_{i}}^{re} \rightarrow \ldots \rightarrow \mathbf{h}_{l, v_{i}}^{*} \rightarrow \ldots \rightarrow \mathbf{h}_{k, v_{i}}^{*},
\end{equation}
where $\mathbf{h}_{l, v_{i}}^{*}$ denotes the representation of item $v_{i}$ which is the output of layer $l$, $l \in (2, k)$, $k$ is the hyper-parameter, and $*$ indicates either $adj$ or $re$.

To fully utilize all features captured by all layers, we apply dense connections \cite{huang2017densely} in our work. The input of each layer consists of the output representations of all previous layers. More specifically, the input of the $l$-th layer is $[\mathbf{h}_{0, v_{i}}^{adj} \| \mathbf{h}_{1, v_{i}}^{re} \| \cdots \| \mathbf{h}_{l-1, v_{i}}^{*}]$. This allows the higher layers to utilize not only the features through their previous layer, but also the low-level features at lower layers. For each item $v_{i}$, we obtain its representations $\mathbf{h}_{v_{i}}^{\prime} \in \mathbb{R}^{d}$ by stacking multiple AIL and RIL.

\subsection{Session Representation Learning Layer}
Through the three layers mentioned above, given session $s=[v_{1}, v_{2}, \ldots, v_{n}]$, we can obtain all item representations in it, i.e.,  $\mathbf{H}=[\mathbf{h}^{'}_{v_{1}}, \mathbf{h}^{'}_{v_{2}}, \ldots, \mathbf{h}^{'}_{v_{n}}]$. Then, we can generate session representation with the representations of items in it.

To reflect the different importance of different positions in the session sequence to the target item, we utilize a learnable position embedding matrix $\mathbf{P}=\left[\mathbf{p}_{1}, \mathbf{p}_{2}, \ldots, \mathbf{p}_{n}\right]$, where $\mathbf{p}_{i} \in \mathbb{R}^{d}$ is a position vector for specific position $i$ and $n$ is the length of the session. 
We combine the position information with item representations through concatenation and non-linear transformation:
\begin{equation}
\small
\begin{split}
\mathrm{z}_{i}=\tanh \left(\mathbf{W}_{2}\left[\mathbf{h}_{v_{i}}^{\prime} \| \mathbf{p}_{n-i+1}\right]+\mathbf{b}_{2}\right), \quad
\mathbf{s}^{\prime} &=\frac{1}{n} \sum_{i=1}^{n} \mathbf{h}_{v_{i}}^{\prime} ,
\end{split}
\end{equation}
where parameters $\mathbf{W}_{2} \in \mathbb{R}^{d \times 2 d}$ and $\mathbf{b}_{2} \in \mathbb{R}^{d}$ are trainable parameters, $\mathbf{s}^{\prime}$ is the session information computed as the average of representations of items in the session.

Next, we adopt a soft-attention mechanism to learn the contribution of item $v_{i}$ to the next prediction, and then we can obtain the session representation by linearly combining the item representations: 
\begin{equation}
\small
\begin{split}
\beta_{i} =\mathbf{q}^{\top} \sigma\left(\mathbf{W}_{3} \mathbf{z}_{i}+\mathbf{W}_{4} \mathbf{s}^{\prime}+\mathbf{b}_{3}\right), \quad
\mathbf{s}=\sum_{i=1}^{l} \beta_{i} \mathbf{h}_{v_{i}}^{\prime},
\end{split}
\end{equation}
where $\mathbf{W}_{3}, \mathbf{W}_{4} \in \mathbb{R}^{d \times d}$ and $\mathbf{q}, \mathbf{b}_{3} \in \mathbb{R}^{d}$ are learnable parameters.

\subsection{Prediction Layer}
We first utilize dot product and then apply softmax function to predict the click probability $\hat{\mathbf{y}}$ for the item $v_i$ :
\begin{equation}
\small
\hat{\mathbf{y}}_{i}=\operatorname{Softmax}\left(\mathbf{s}^{\top} \mathbf{h}_{v_{i}}\right),
\end{equation}
where $\hat{\mathbf{y}}_{i} \in \hat{\mathbf{y}}$ denotes the probability of item $v_{i}$ to be the true next item.
The loss function is defined as the cross-entropy of the prediction results.

\section{Experiments and Analysis}

\subsection{Experimental Settings}

\subsubsection{Datasets.}

We select two datasets from the Amazon dataset\footnote{https://nijianmo.github.io/amazon/index.html} for our experiments: \emph{Pet Supplies} and \emph{Movies and TV}. The datasets contain purchase history of users and users' reviews for products. Following a common manner, we remove items appearing less than 5 times. Following \cite{song2019session}, we split user's purchase behaviors into week-long sessions. To evaluate our method more comprehensively, we prepare two versions for each dataset. The first version (Case 1) keeps all sessions with more than 1 item~\cite{li2017neural,wang2020global,wu2019session} while the second version (Case 2) keeps sessions with more than 5 items~\cite{wu2019session}. Obviously, Case 2 is a subset of Case 1 which keeps long sessions only. We set the sessions of last year as the test data, and the remaining sessions for training. Then, we adopt sequence splitting preprocessing method which is commonly adopted in SBR. For an input session $[v_{1}^{s}, v_{2}^{s}, \ldots, v_{n}^{s}]$, we generate multiple input sequence-label pairs, i.e., $([v_{1}^{s}], v_{2}^{s}),([v_{1}^{s}, v_{2}^{s}], v_{3}^{s})$, \ldots, $([v_{1}^{s}, v_{2}^{s}, \ldots, v_{n-1}^{s}], v_{n}^{s})$.

After the preprocessing, the statistic of datasets are summarized in Table \ref{tab:statistic}.
\begin{table}[H]
\caption{Statistic details of datasets.}
\label{tab:statistic}
\scriptsize
\begin{center}
\begin{tabular}{c|c|c}
\hline Dataset & Pet Supplies & Movies and TV \\
\hline \# Interactions & 1416,391 & 2114,752 \\
\hline \# Train sessions & 894,646 & 586,696 \\
\hline \# Test sessions & 99,024 & 20,823 \\
\hline \# All the items & 30,694 & 46,341 \\
\hline \# Average session length & 3.35 & 3.48 \\
\hline
\end{tabular}
\end{center}
\end{table}

\subsubsection{Evaluation Metrics and Baselines.}
Following \cite{wang2020global,wu2019session}, we adopt P@K (Precision) and MRR@K (Mean Reciprocal Rank) as evaluation metrics. 
We compare RI-GNN with the following representative methods, including {S-POP} \cite{jannach2017sknn}, {S-KNN} \cite{jannach2017sknn}, {GRU4Rec} \cite{HidasiKBT16}, {NARM} \cite{li2017neural}, {STAMP} \cite{liu2018stamp}, {BERT4Rec} \cite{sun2019bert4rec}, {SR-GNN} \cite{wu2019session}, {GCE-GNN} \cite{wang2020global} and {DHCN} \cite{xia2021dhcn}. (1) \textbf{S-POP} recommends the top-\emph{N} frequent items in the current session. (2) \textbf{S-KNN} \cite{jannach2017sknn} is a session-based k-nearest-neighbors approach. (3) \textbf{GRU4Rec} \cite{HidasiKBT16} applies RNN to SBR, which adopts GRU to model short-term user preferences. (4) \textbf{NARM} \cite{li2017neural} is a RNN-based model that utilizes attention mechanism to capture user's main intents and sequential behavior. (5) \textbf{STAMP} \cite{liu2018stamp} employs MLP and attention mechanism to capture the long-term and short-term interests of users, respectively. (6) \textbf{BERT4Rec} \cite{sun2019bert4rec} introduces the Bert model to capture item transitions for the sequential recommendation. (7) \textbf{SR-GNN} \cite{wu2019session} converts each of session to a directed graphs and updates item embeddings by gated GNN. (8) \textbf{GCE-GNN} \cite{wang2020global} captures item-transitions from all sessions, not just a single session. (9) \textbf{DHCN} \cite{xia2021dhcn} captures the complex high-order information among items by modeling session-based data as a hypergraph.

\subsubsection{Hyper-parameters Settings.}
Following previous studies \cite{li2017neural,wang2020global,wu2019session}, the dimension of node embedding is 100, the size of mini-batch is 100, and the $L_{2}$ regularization is $10^{-5}$ for all models. For RI-GNN, we use the Adam optimizer with the initial learning rate 0.001. The dropout ratio of session graph is 0.2. Moreover, the number of topics are empirically set to 24 and 20 on \emph{Pet Supplies} and \emph{Movies and TV} dataset, respectively.

\subsubsection{Overall Performance.}

\begin{table*}
\centering
\scriptsize
\caption{\footnotesize Experimental Results on Sessions with More than 1 Item.}
\label{tab:overallperformance}
\resizebox{0.8\textwidth}{!}{
\begin{tabular}{c|cccc|cccc}
\hline
Dataset     & \multicolumn{4}{c|}{Pet Supplies}                                 & \multicolumn{4}{c}{Movies and TV}                                 \\ \hline
Metrics     & P@10           & P@20           & MRR@10         & MRR@20         & P@10           & P@20           & MRR@10         & MRR@20         \\ \hline
S-POP       & 6.52           & 6.52           & 5.25           & 5.26           & 1.83           & 1.84           & 1.59           & 1.59           \\
S-KNN        & 21.86         & 24.30           & 12.22          & 12.39          & 20.33          & 23.46          & 8.40            & 8.62           \\ \hline
GRU4Rec     & 8.79           & 9.84           & 6.26           & 6.34           & 8.32           & 9.47           & 5.64           & 5.72           \\
NARM        & 24.48          & 27.68          & 19.48          & 19.70           & 22.40          & 25.40          & 17.05          & 17.26          \\
STAMP       & 21.50          & 24.66          & 16.41          & 16.60           & 20.59          & 23.96          & 14.69          & 14.92          \\
BERT4Rec    & 24.18          & 27.40           & 19.29          & 19.51         & 23.21          & 26.46          & 17.75          & 17.97          \\ \hline
SR-GNN      & 24.64          & 27.81          & {\ul 19.53}    & {\ul 19.75}    & 23.69          & 26.74          & {\ul 18.24}    & {\ul 18.45}    \\
GCE-GNN     & {\ul 24.73}    & {\ul 28.28}    & 18.16          & 18.41          & {\ul 24.25}    & {\ul 27.64}    & 17.02          & 17.25          \\
DHCN        & 24.23          & 27.45          & 17.80           & 18.02          & 23.33         & 26.49         & 15.94          & 16.16          \\ \hline
RI-GNN       & $\mathbf{25.43}^{*}$ & $\mathbf{28.88}^{*}$ & $\mathbf{19.93}^{*}$ & $\mathbf{20.15}^{*}$ & $\mathbf{24.69}^{*}$ & $\mathbf{27.94}^{*}$ & $\mathbf{18.93}^{*}$ & $\mathbf{19.14}^{*}$ \\
Improv.(\%) & 2.83           & 2.12           & 2.05           & 2.02           & 1.81           & 1.09           & 3.78           & 3.74           \\ \hline
\end{tabular}}
 \begin{tablenotes}
        \scriptsize 
         \item[1] $^{1}$ {The best results of each column are highlighted in boldface, the suboptimal one is underlined, the improvements are calculated by using the difference between the performance of our proposed RI-GNN and the best baseline, and * denotes the significant difference for t-test.}
 \end{tablenotes}
\end{table*}

\begin{table*}[htbp]
\centering
\caption{\footnotesize Experimental Results on Sessions with More than 5 Items.}
\label{tab:5-length sessions}
\resizebox{0.8\textwidth}{!}{
\begin{tabular}{c|cccc|cccc}
\hline
Dataset     & \multicolumn{4}{c|}{Pet Supplies}                                 & \multicolumn{4}{c}{Movies and TV}                                 \\ \hline
Metrics     & P@10           & P@20           & MRR@10         & MRR@20         & P@10           & P@20           & MRR@10         & MRR@20         \\ \hline
NARM        & 26.07          & 30.34          & 19.33          & 19.62          & 20.98          & 24.73          & 14.27          & 14.53          \\
STAMP       & 24.44          & 28.06          & 18.28          & 18.53          & 21.27          & 25.40          & 14.49          & 14.77          \\
BERT4Rec    & 25.85          & 30.04          & {\ul 19.49}    & {\ul 19.77}    & 22.24          & 26.43          & 15.87          & 16.15          \\ \hline
SR-GNN      & {\ul 26.29}    & {\ul 30.49}    & 19.31          & 19.60          & 22.73          & 26.57          & {\ul 16.11}    & {\ul 16.37}    \\
GCE-GNN     & 25.95          & 30.36          & 17.51          & 17.81          & {\ul 23.18}    & {\ul 27.51}    & 14.39          & 14.71          \\
DHCN        & 24.96          & 29.20          & 17.25          & 17.54          & 21.83          & 25.71          & 13.01          & 13.27          \\ \hline
RI-GNN      & $\mathbf{27.72}^{*}$ & $\mathbf{32.11}^{*}$ & $\mathbf{20.26}^{*}$ & $\mathbf{20.57}^{*}$ & $\mathbf{24.15}^{*}$ & $\mathbf{28.30}^{*}$ & $\mathbf{16.99}^{*}$ & $\mathbf{17.27}^{*}$ \\
Improv.(\%) & 5.44           & 5.31           & 3.95           & 4.05           & 4.19           & 2.87           & 5.46           & 5.50           \\ \hline
\end{tabular}}
\end{table*}

The experimental results of overall performance on sessions of different lengths are reported in Table \ref{tab:overallperformance} and Table \ref{tab:5-length sessions} respectively, according to the tables, we can draw the following conclusions:

\textbf{Case 1} (Performance on sessions with more than 1 item): (1) Traditional methods (i.e., S-POP, S-KNN) show a significant inferiority to neural methods, except for GRU4Rec. This demonstrates that neural networks can learn more sophisticated features than the traditional methods. (2) Neural network based methods (i.e., GRU4Rec, NARM, STAMP, BERT4Rec) usually have better performance for SBR. GRU4Rec shows worse performance, which is probably because it strictly defines a session as a sequence, and does not consider the diversity of user preferences. The other neural methods (i.e., NARM, STAMP, BERT4Rec) outperform GRU4Rec significantly. Among them, NARM combines RNN and attention mechanism, STAMP and BERT4Rec are completely based on attention mechanism. This result demonstrates that attention-based methods are also an effective way besides the RNN-based methods. (3) Among all the baseline methods, the GNN-based methods (i.e. SR-GNN, GCE-GNN and DHCN) demonstrate the superiority over the others. They convert each session into a graph and model item dependencies with GNN. This indicates that GNN-based models would be more effective than RNN-based and attention-based models when capturing the complex dependencies between items in SBR. (4) It is obvious that our proposed RI-GNN model outperforms all the baselines on all datasets. Compared with the GNN-based methods, the superiority of RI-GNN may due to that it is equipped with RIL and review information, which can model item dependencies more accurately.

\textbf{Case 2} (Performance on sessions with more than 5 items): Table \ref{tab:5-length sessions} shows that our proposed RI-GNN model can outperform the state-of-the-art SBR methods by 2.87\% to 5.50\% on different metrics. Compared with the result in Table \ref{tab:overallperformance}, RI-GNN outperforms state-of-the-art SBR methods with a larger margin on long sessions. This demonstrates the superiority of RI-GNN is more outstanding for the long sessions. The reason may be that there exist more dependencies between non-adjacent items within the long sessions. This further verifies our hypothesis.

\begin{figure}[t]
\centering
\includegraphics[width=\linewidth]{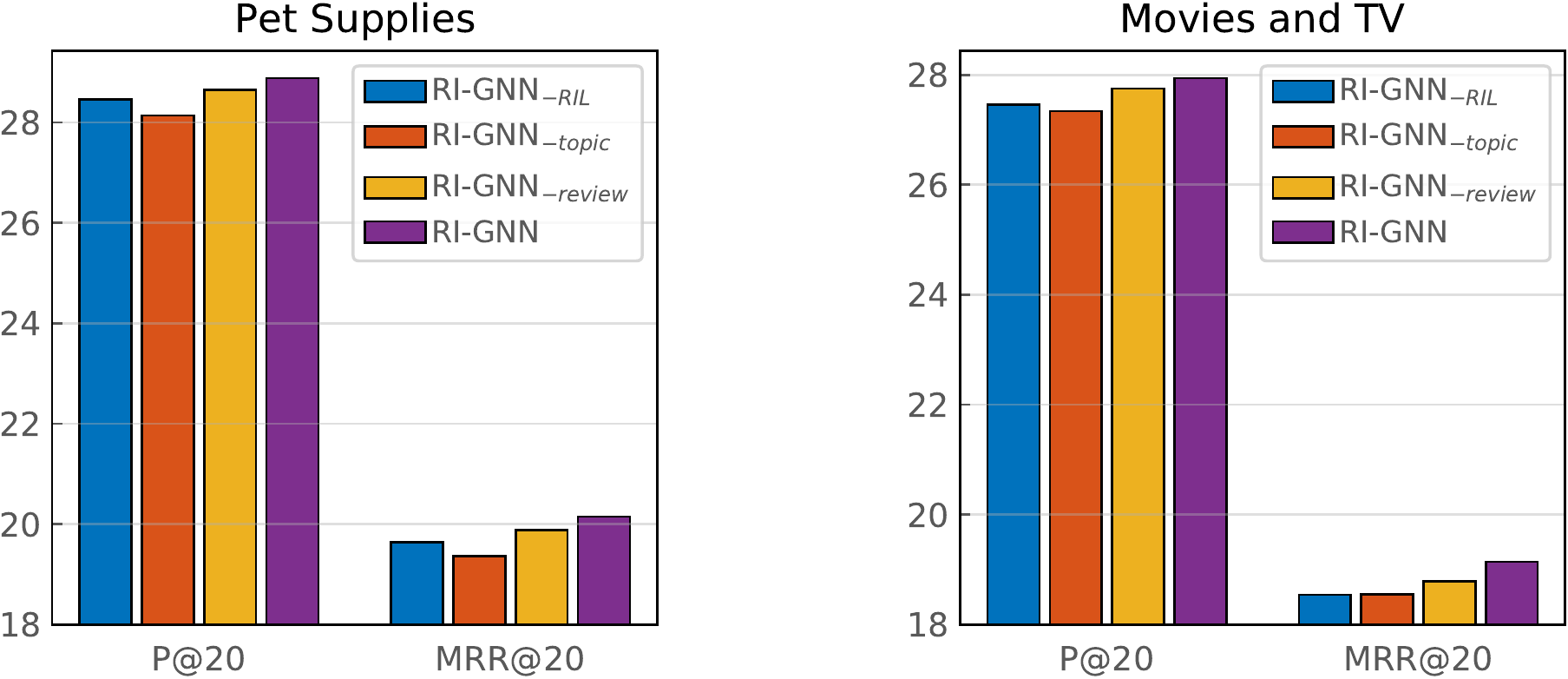}
\caption{Comparison of Ablation Variants.}
\label{fig:ablation study}
\end{figure}

\subsubsection{Ablation Study.}

To investigate the effectiveness of RIL component and review information for RI-GNN, we implement three variants of RI-GNN, denoted as RI-GNN$_{-RIL}$, RI-GNN$_{-topic}$, and RI-GNN$_{-review}$, by removing RIL, topic and review respectively. The comparison of them is shown in Fig. \ref{fig:ablation study}. It is clear that all variants are inferior to the standard RI-GNN, which demonstrates that RIL, topics and review information are critical and necessary for the success of RI-GNN model.

\section{Conclusion}
In this paper, we rethink adjacent dependencies in SBR, argue that adjacent items in a session are not always dependent and non-adjacent items are not necessarily independent. Accordingly, we propose a novel review-refined inter-item graph neural network (RI-GNN), which leverages reviews to reduce the false dependencies between adjacent but actually independent items and capture true dependencies between non-adjacent but dependent items. Empirical evaluations on two real-world datasets demonstrate the superiority of RI-GNN over the state-of-the-art methods. In the future, we will explore more powerful methods to refine the dependencies within sessions.

\noindent\textbf{Acknowledgment.} Wenpeng Lu is the corresponding author. The research work is partly supported by National Natural Science Foundation of China under Grant No.11901325 and No.61502259, National Key R\&D Program of China under Grant No.2018YFC0831700, Key Program of Science and Technology of Shandong Province under Grant No.2020CXGC010901 and No.2019JZZY020124, and Natural Science Foundation of Shandong Province under Grant ZR2021MF079.

\bibliographystyle{ACM-Reference-Format}
\bibliography{RI-GNN}

\end{document}